\documentclass{IET}
\begin{document}

\title{Analytic Model for Network Resource Management between ISPs and Users}

\author[1,*]{Hossein Lolaee}
\affil{ECE Department,  University of Tehran, North Kargar St., Tehran, Iran}

\author[1]{Mohammad Ali Akhaee}

\affil[*]{hossein\_lolaee@ut.ac.ir}

\abstract{Fixed Communication Provider (FCP) is a consortium of Internet Service Providers (ISPs)  which users can switch  easily and freely between their ISPs. In order to increase the QoS of the ISPs, we propose a two class service model as the following. 
	ISPs divide their available bandwidth into two parts to provide their end users with optimal services. One dedicated to primary users, and the other for secondary users. Primary users are those who pay more and thus, expect dedicated bandwidth that is always available. Secondary services are provided by ISPs for the other users who cannot afford the dedicated bandwidth. In this study, by defining the utility functions for both user types, we aim at dividing the ISP bandwidth between these two services such that the utility function of the users is maximized. Since the primary users do not always use the maximum bandwidth, an algorithm is proposed for ISPs to estimate the primary users' required bandwidth in each time segment based on the previous segments. Based on this estimate, the expected bandwidth is dedicated to the primary users, and the remaining bandwidth is devoted to the secondary users to improve the Quality of Service (QoS). On the other hand, an ISP is penalized if it fails to provide the primary users with the required bandwidth mentioned in their contract. Therefore, there exists a trade-off between the conservative estimation of the primary users' rate, QoS of the secondary users and the achieved utility of the ISPs.
	\\ \\
		\textbf{keywords}:
		Network analysis,  Network management, Quality of service (QoS).}
\maketitle

\section{Introduction}
\label{sec:intro}
Rapid development of technology and the need for online information has made the Internet an important necessity for our daily lives. Internet users form a wide range of individuals with different needs. Some people use the Internet only for their businesses, while others may need it for their personal and social communications. Therefore, different individuals have different expectations from Internet services. For example, QoS is important to users of the first category as low QoS can cause them financial losses. However, QoS may not be as important for users in the second group. 

Although technological developments have enabled ISPs to provide different services for different types of users, this idea is not yet  realized \cite{1}. Currently, ISPs only offer services with predefined user's maximum rate or traffic \cite{2}. Recently, some ISPs have suggested services based on the packets prioritization \cite{3}, but this approach contradicts the network neutrality regulation. It is mentioned in the network neutrality regulation that the government and ISPs must treat the Internet data equally and must not differentiate between Internet data based on their user type, content, website and application. Supplementary information in this regard can be found in \cite{4,5}.

Several approaches are proposed in the literature to provide more efficient Internet services \cite{6,Etkin_2007}. A common approach is that the Internet services are divided into different classes. In  Paris Metro Pricing  (PMP) method proposed by Odlyzko, different QoS's are achieved by two to four service classes with different pricings \cite{9}. 
Shetty \textit{et. al.} have proposed a two-class service for the Internet based on the PMP model \cite{2}. In this model, the ISP bandwidth is dedicated to two services named primary and secondary services, based on   maximizing  three scenarios:
The utility of  a class of  users,  the utility of  only users  and the total utility of the ISP and users.
In \cite{Shetty_2009}, an optimal pricing scheme was obtained for a monopolistic market and the authors analyzed
the economic impacts  on the providers' revenue, while in \cite{Ren_2012}, the authors  studied the dynamics of the users' demand in a monopolistic manner. In 
\cite{Lingjie_Duan_2014}, they considered a simplified framework with only one primary user to increase his performance while satisfying the secondary users' requirements under incomplete channel state information (CSI) in the cognitive radio.
None of these papers, considered  cooperative strategy   to manage the network resources between multiple primary and secondary users. 
 Moreover, we will show by  considering the dynamics of users' demand,  that both ISPs and users can be even more satisfied.

In this paper, we propose a two class service model in order to analayze the interaction between ISPs and users who are in the same FCP. 
 Considering multiple primary and secondary users, we provide a novel analytic model  between ISPs and users.
The core idea is similar to the cooperative spectrum sharing in cognitive radio literature (e.g. \cite{Gao_2013,Shi_2014,Sharma_2016}) where secondary users collaborate with primary users  to use their bandwidth.
 ISP bandwidth is divided into two parts. In the first part, the bandwidth is dedicated to the primary users who expect a dedicated bandwidth due to their higher payment. In the second part, there are secondary users who expect the minimum QoS declared in the contract between them and the ISP.  
We define the utility functions for both user types, then divide the ISP bandwidth between these two services such that the utility function is maximized. 
 Since the primary users do not always use the total perchased bandwidth, a method is proposed for ISPs to estimate the primary users' required bandwidth in each time segment based on the previous segments. We attempt to demonstrate the benefits of this strategy for both ISPs and users. 

This paper is organized as follows. In section \ref{sec:prpsd}, ISP services for the users are explained, and analytic models for both primary and secondary users are presented. In section \ref{sec:analysis}, the presented model is analyzed  and the results for both users are given. Finally in section \ref{sec:cncl}, the paper is concluded.

\section{Proposed Game Theory Model}
\label{sec:prpsd}
\subsection{Problem Definition}
\label{ssec:def}

In order to provide better and more economic services, ISPs divide their end-users into two groups, primary users and secondary users. The primary users receive high quality services in return of higher payment. In fact, these users buy a dedicated bandwidth and expect it to be always available to them by the  ISP. Another service type is given to secondary users who do not expect very high QoS or cannot afford buying a dedicated bandwidth.
Secondary users expect minimum QoS. The available bandwidth for these users is variable in time. For instance, the available bandwidth approaches its minimum values during the daytime when there is more demand for Internet services, while the bandwidth increases in the last hours of the day and midnight, due to the reduction in the demand. 

\begin{figure}
\centering
\includegraphics[width=0.7\linewidth]{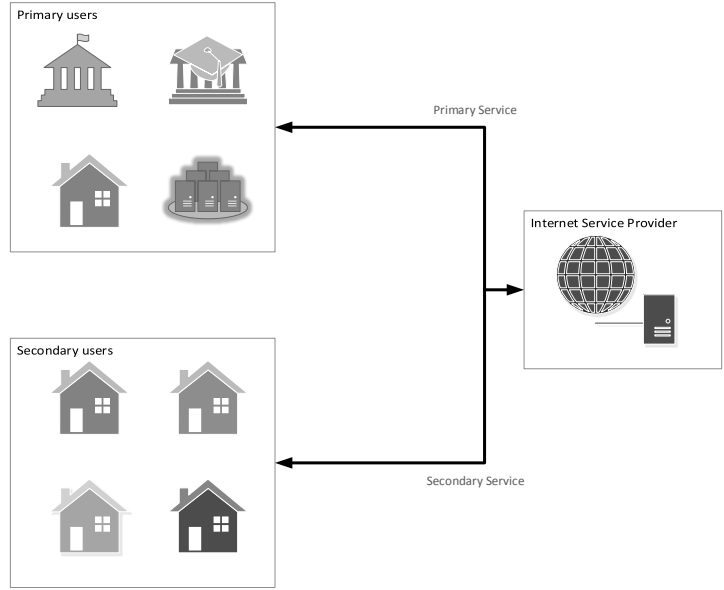}
\caption{A model for  ISP services to the users.}
\label{fig:diagram}
\end{figure}

Science and research centers, universities, government organizations, banks, large companies and even home users who need dedicated bandwidths are some examples of primary users. Secondary users are typically home users for whom buying a constant bandwidth is not economical. This classification is shown in Fig. \ref{fig:diagram}.

As it is already mentioned, each ISP provides two types of services. With the first type of service, user enjoys a trusted service in return for paying more money. With the second service type, a user accepts the minimum quality offered by the ISP, in return for paying less money. In this study, we as the regulatory organization, aim at distributing the bandwidth available to each ISP between the primary and secondary users in a way that both sides are fairly satisfied, in addition to determining the price of both service types. 

\subsection{Problem Modeling}
\label{ssec:model}

Assume that there exist $M$ (constant) competing ISPs, each of which providing both service types to their customers. The super script $m=1,\dots ,M$ denotes the variable related to the $m$'th ISP. Assume that there exists a total number of $N$ end users, with $N$ being a large number. Each ISP has purchased the bandwidth $C^m>0$ for the cost of $\tau >0$ per bandwidth unit. The ISPs' investment on purchasing the bandwidth is irreversible and cannot be changed. The available bandwidth $C^m$ is divided into two parts $C_p^m$ and $C_s^m$, representing the bandwidths dedicated to the primary and secondary services, respectively $(C^m=C_p^m+C_s^m)$. $Z^m$ denotes the number of users who have received their required service from the $m$'th ISP, and equals the total number of the primary users ($Z_p^m$) and secondary users ($Z_s^m$). In the following, we indicate the analytic model  between the secondary users and ISPs at first, and then consider the same problem for the primary users. 

\subsubsection{Secondary Service}
\label{sssec:secondary}

In order to present its secondary service, each ISP announces $q_{s,min}^m$ and $p_s^m$ to the secondary users. The subscript $s$ denotes the secondary users. $q_{s,min}^m$ determines the lowest QoS offered by the $m$'th ISP to the secondary users, and $p_s^m$ is the price of such service. $q_s^m$ denotes the quality of the momentary service provided by the $m$'th ISP. It should be noticed that after signing the contract between the ISP and secondary users, these users no longer are charged for any momentary rise in quality higher than  $q_{s,min}^m$. Therefore, the secondary users have the chance of accessing higher bandwidths when the demand is low, without paying any additional costs, while they expect the QoS of $q_{s,min}^m$  according to the contract. Note that the minimum quality of $q_{s,min}^m$ must  always be guaranteed by the ISP. 

When ISPs declared the QoS and prices, $q_{s,min}^m$ and $p_s^m$ must find values in the market with their minimum ratio equal to $E$, which is defined as below:
\begin{equation}
E = \min_i \frac{{p_s^i}}{{q_{s,min}^i}}.
\label{eq:balance}
\end{equation}

Equation (\ref{eq:balance}) is in correspondence with the reality of the price balance in the market. ISPs balance their prices and QoS such that they all come to a consistent quality to price ratio in a balanced situation. In fact, a user switches to a new ISP which offers lower price per bandwidth unit. From a different viewpoint, if ISPs find out that there are other ISPs offering lower prices for the same QoS, they will  either decrease their prices or increase their QoS to avoid loss of their customers and income. QoS perceived by the users connected to the $m$'th ISP is defined as:
\begin{equation}
q_s^m = 1- \frac{{Z_s^m}}{{C_s^m}}.
\label{eq:qossecond}
\end{equation}

This definition is consistent with our mentality about QoS \cite{10}, that is, QoS increases when ISP promotes the bandwidth available to the secondary users while the number of users is constant, or decreases when the number of users  increases while the available bandwidth is constant. In the QoS model, it is assumed that all users affect the QoS in the same way, that is, the traffic generated by different users are equal.
To simplify the problem modeling process, the bandwidth and the number of users are normalized to values between zero and one, and are denoted by $z_s^m$ and $c_s^m$, respectively. $z_s^m=\frac{Z_s^m}{N}$ is the ratio of the end users connected to the $m$'th ISP. $c_s^m=\frac{C_s^m}{N}$ denotes the potential bandwidth available for each end user. Note that such normalization does not affect the definition and equation of the QoS. Thus, (\ref{eq:qossecond}) can be restated as:
\begin{equation}
q_s^m = 1- \frac{{z_s^m}}{{c_s^m}}.
\label{eq:qossecond2}
\end{equation}

Each $i$'th end user is known by a type $\theta_s^i$ which is a random variable at the range of $[0,1]$. $\theta_s^i$ shows the QoS priority for the secondary user. The greater the $\theta_s^i$ value, the more important  the QoS for a secondary user \cite{11}. Now, the utility of the $i$'th user connected to the $m$'th ISP can be calculated as below:
\begin{equation}
U_s^{m,i}=\theta_s^i q_s^m- p_s^m,
\label{eq:Utilitys}
\end{equation}
where $\theta_s^i$ is the user required network QoS in his maximum traffic. For instance, a user who uses Internet for watching television channels has a larger $\theta_s^i$ compared to a user who needs the Internet service only to check his email \cite{2}.
The user distribution function on $\theta_s^i$ is denoted by $p(\theta)$ at the range of $[0,1]$. Hence, the average utility of the secondary user who is connected to the $m$'th ISP can be expressed as:
\begin{equation}
\overline{U}_s^m=\int_0^1 U_s^{m,i} p(\theta) \mathrm{d}\theta.
\label{eq:bUtilitys}
\end{equation}
Normally in the literature, $p(\theta)$ is taken as the uniform distribution \cite{12,13}.

\subsubsection{Primary Service}
\label{sssec:primary}

ISPs announce the primary service price $p_p^m$ per bandwidth unit to primary users. In the   balanced-price market situations, the final price is determined similar to the secondary service case:
\begin{equation}
p_p=\min_m p_p^m.
\label{eq:balancep}
\end{equation}

In this way, the user must pay more for a higher bandwidth. In comparison with secondary service,  each user after connecting to the network, expects to constantly access the bandwidth he has purchased  from the ISP. In order to better manage the network resources, ISPs do not always provide primary users with the total purchased bandwidth, because sometimes the user is not connected to the network at all and it is not rational to dedicate to him a certain bandwidth in such cases. In fact, the bandwidth bought by the user must be assigned based on the user's demand.  

For this purpose, ISPs break the time into timeslots. They decrease the assigned bandwidth to the primary user if his used traffic has not been significant in a few latest timeslots. Otherwise, if the traffic is increasing, the total bandwidth bought by the primary user is dedicated to him. At the $k$'th timeslot, ISP dedicates the rate of $g(r_p^i,\alpha_k)$ to the user by adjusting a parameter named $\alpha_k$, where $r_p^i$ denotes the bandwidth purchased by the $i$'th user from the ISP, and $g(r_p^i,\alpha_k)$ is named the limiter function due to its nature. If the ISP fails to dedicate the demanded rate of the user in each timeslot, it must pay the penalty of $\lambda$ per the rate unit, proportional to the amount of the decrease in the bandwidth demanded by the user. 

The bandwidth requested from the network by the $i$'th user at the $k$'th timeslot is represented by $r_p^{i,k}$. Moreover, each primary user has a type $\theta_p^i$ that is defined in a way similar to the secondary user. Now the utility function for the $i$'th primary user can be calculated as:
\begin{equation}
U_p^i=\theta_p^i g(r_p^i,\alpha_k) -r_p^i p_p + \lambda I\Big(r_p^{i,k}-g(r_p^i,\alpha_k) \Big) 
\label{eq:Utiltyp}
\end{equation}
in which $I(x)$ is defined as:
\begin{equation}
I(x)= \left\{
\begin{array}{ll}
0 & \text{ if} \; x < 0\\
x & \text{ if} \; x >=0
\end{array} \right..
\label{eq:Ix}
\end{equation}

In order to simplify the problem solving procedure, $p^i$ and $r_p^{i,k}$ are assumed to be normalized by $C_p^m$. Then, we need to find the function $g(x,\alpha_k)$ such that $0<x,\alpha_k,g(x,\alpha_k)<1$, where $\alpha_k$ is an adjusting parameter used by ISP to change the bandwidth dedicated to the user at the $k$'th timeslot. The ISP dedicates the total sold bandwidth when $\alpha_k$ takes its maximum value, and provides the user a minimum bandwidth when $\alpha_k$ finds its minimum. Furthermore, the limiter function must be designed to change with respect to $\alpha_k$ in  smaller $\alpha_k$ values and it is greater than those in  large $\alpha_k$ values. The reason is that the user's traffic is usually burst and for the ISP to manage to follow the momentary user's rate, the limiter function must be more sensitive in  smaller $\alpha_k$ values and the rate must change considerably in such situations. This can be concluded in the following equations:
\begin{equation}
g(x,\alpha_k): \left\{
\begin{array}{l}
g(x,1)=x\\
g(x,0)=r_{min} \\
\frac{{\partial g(x,\alpha_k)}}{{\partial \alpha_k}} \Big\vert_{\alpha_k \approx 0} > \frac{{\partial g(x,\alpha_k)}}{{\partial \alpha_k}}  \Big\vert_{\alpha_k \approx 1} 
\end{array} \right..
\label{eq:Gcond}
\end{equation}
in which $r_{min}$ is the minimum bandwidth ISP provides for each primary users.
Many functions can be suggested to satisfy these three constraints. In order to have a tractable solution, the definition of the gamma function is exploited. The proposed function is defined in the equation below:
\begin{equation}
g(x,\alpha_k)=(x-r_{min})  (\alpha_k)^{\gamma}+r_{min},
\label{eq:Gdefin}
\end{equation}
where $\gamma$ is the adjusting parameter of the limiter function at the range of $[0,1]$.  For a  better realization, for $x=0.9$, $r_{min}=0.2$, and different $\gamma$ values, $g(x,\alpha_k)$ is drawn in Fig. \ref{fig:gx}.
$\alpha_k$ is an adjusting parameter determining the user's traffic in the last timeslots, and can be adjusted as below:
\begin{figure}
	\centering
	\includegraphics[width=0.7\linewidth]{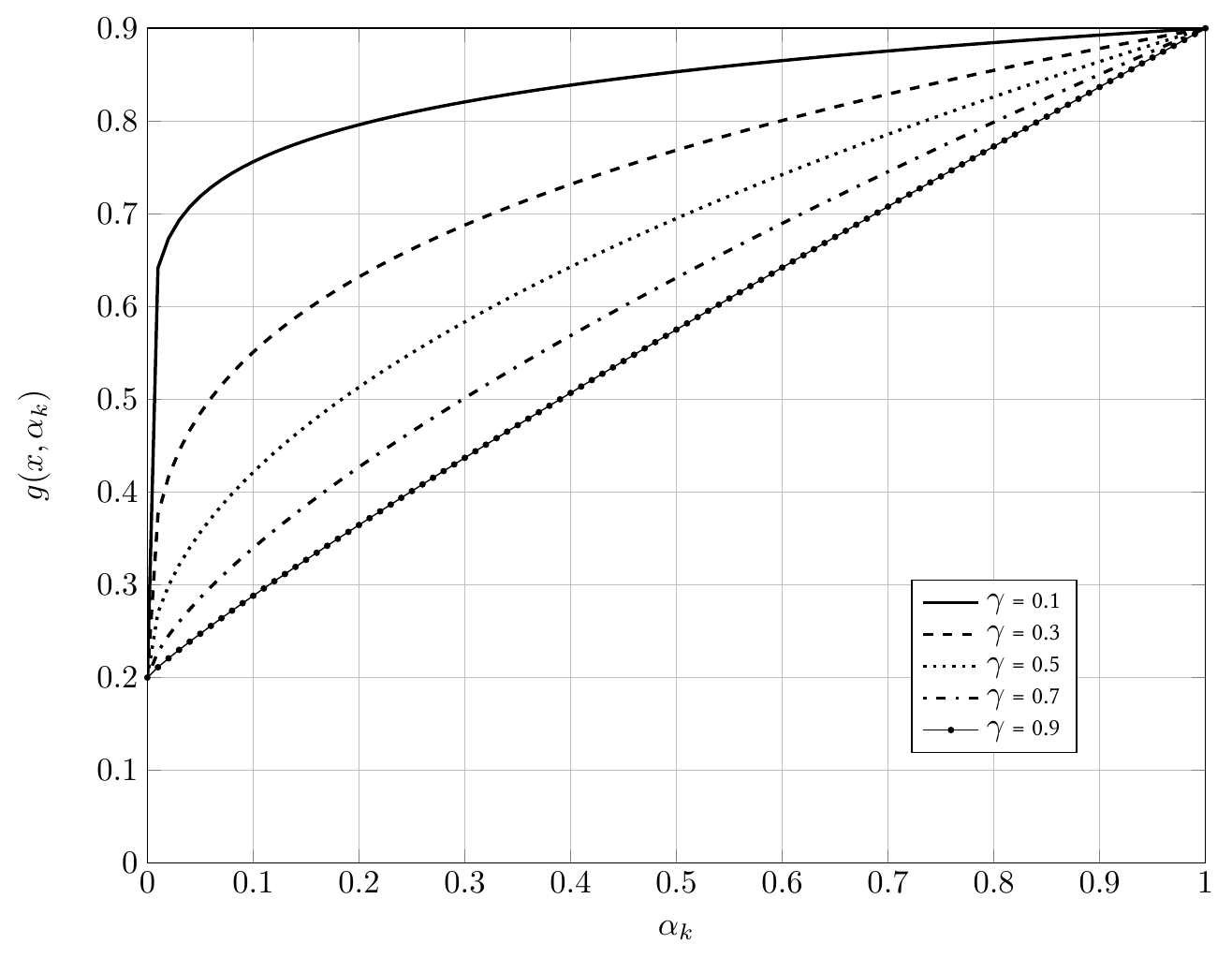}
	\caption{$g(x,\alpha_k)$ function.}
	\label{fig:gx}
\end{figure}
\begin{equation}
\alpha_k =\frac{\displaystyle\sum_{k'=k-t}^{k-1}\omega_{k'} \mathbf{1}\Big(r_p^{i,k'}-g(r_p^i,\alpha_{k'})\Big)}{t},
\label{eq:alphak}
\end{equation}
where the function $\textbf{1}(x)$ is defined as:
\begin{equation}
\mathbf{1}(x)= \left\{
\begin{array}{ll}
0 & \text{ if}x < 0\\
1 & \text{ if}x >=0
\end{array} \right.,
\label{eq:1x}
\end{equation}
and the variable $t$ denotes the number of timeslots involved in determining $\alpha_k$ . $\omega_{k'}$ coefficients must be chosen to minimize the difference between the $g(x, \alpha_k )$ function and the real value at the $k$'th timeslot. According to Fig. \ref{fig:gx}, the $g(x, \alpha_k )$ function is more nonlinear for the smaller $\gamma$ values and can better trace the increasing burst traffics, although the tracing is not satisfactory in case of the decrease in the burst traffics. 

\section{Model Analysis}
\label{sec:analysis}
In this section, the problem model defined in section \ref{sec:prpsd} would be analyzed. We begin with the secondary service for this purpose, and then proceed with the analysis for the primary service.

\subsection{Secondary Service Analysis}
\label{ssec:sa}

According to section \ref{sssec:secondary}, utility function of the $i$'th user connected to the $m$'th ISP equals $U_s^{m,i}=\theta_s^iq_s^m-p_s^m$. Obviously, the user does not use the secondary service of the ISP if his utility is negative. Thus, the secondary user must enjoy the positive $U_s^i$, and their type must be in the form of $\theta_s^i\geq \frac{p_s^m}{q_s^m}$. Note that $q_s^m$ denotes the momentary quality. To obtain a more conservative range for $\theta_s^i$, $q_s^m$ is replaced with its minimum value, $q_{s,min}^m$. Then, we have:
\begin{equation}
\theta_s^i \geq \frac{p_s^m}{q_{s,min}^m} = E.
\label{eq:limt1}
\end{equation}
$E$ is defined in (\ref{eq:balance}). Therefore, the secondary users' type falls within the range of $[E,1]$. Referring to the discussion in section \ref{sssec:secondary}, the type of the secondary users is assumed to be uniformly distributed over the range of $[0,1]$. Hence, the number of the secondary users connected to the $m$'th ISP equals $z_s^m=1-E$.
The $i$'th user's income is defined as $S_s^i=\theta_s^iq_s^m$. Similar to the definition of the average utility in (\ref{eq:bUtilitys}), the average income is defined and $p(\theta)$ is set to the uniform distribution. Since the users' type is in the range of $[E,1]$, we have:
\begin{equation}
\overline{S}_s^m = \int_E^1 S_s^{m} p(\theta) \mathrm{d}\theta_s^i  = \int_E^1 \theta_s^i q_s^m \mathrm{d}\theta_s^i  
= 1/2 \times q_s^m (\theta_s^i)^2\Big \vert_E^1=1/2 \times q_s^m (1-E^2).
\label{eq:Sint}
\end{equation}
Substituting (\ref{eq:qossecond2}) and $z_s^m=1-E$ in (\ref{eq:Sint}) yields:
\begin{equation}
\overline{S}_s^m = 1/2 \times  (1- \frac{{z_s^m}}{{c_s^m}}) \Big(1-(1-z_s^m)^2\Big) 
= \frac{{1}}{{2 c_s^m}} \times  (c_s^m- z_s^m) \Big(1-(1-z_s^m)^2\Big).
\label{eq:Sint2}
\end{equation}
Now, we aim at finding the number of the secondary users with their average income being maximized. Thus, we take the derivative of (\ref{eq:Sint2}) with respect to $z_s^m$:
\begin{equation}
\frac{\partial 	\overline{S}_s^m}{\partial z_s^m}=0 \Longrightarrow -1+(1-z_s^m)^2 + 2\times (c_s^m - z_s^m) (1-z_s^m)=0.
\label{eq:Derivs}
\end{equation}
In  (\ref{eq:Derivs}) $1-z_s^m$ is replaced with E for the ease of calculations; Thus, we have:
\begin{equation}
-1+E^2 + 2\times c_s^m E- 2\times E(1-E)=0
\Longrightarrow 3E^2+2(c_s^m-1)E-1=0.
\label{eq:Eeq}
\end{equation}
Solving    (\ref{eq:Eeq}) yields the number of the secondary users with the maximum average income:
\begin{equation}
z_s^m = 1-\frac{1-c_s^m + \sqrt{(1-c_s^m)^2+3}}{3}.
\label{eq:zsequi}
\end{equation}
Note that since the other root of (\ref{eq:Eeq}) is negative and thus unacceptable, it is not included in (\ref{eq:zsequi}).

In the following, we want to calculate the ratio of the bandwidth available to the ISPs that is dedicated to secondary users. The $x$ ratio is defined as $c_p^m=xc^m$ for $x\in [0,1]$, $c_p^m=\frac{C_p^m}{N}$ and $c^m=\frac{C^m}{N}$. Regarding the fact that $c^m=c_p^m+c_s^m$, we have $c_s^m=(1-x)c^m$. Using (\ref{eq:Eeq}) the bandwidth $c_s^m$ can be found as:
\begin{equation}
c_s^m = \frac{2E(1-E)+1-E^2}{2E}=1-\frac{3}{2}E+\frac{1}{2E}.
\label{eq:csm}
\end{equation}
Thus, we have:
\begin{equation}
x = 1+\frac{1}{2c^m}\Big(3E-\frac{1}{E}-1\Big).   
\label{eq:xeq}
\end{equation}
Substituting $E=\frac{p_s^m}{q_{s,min}^m}$ in (\ref{eq:xeq}) it can be restated as:
\begin{equation}
x=1+\frac{1}{2c^m}\Big(3\frac{p_s^m}{q_{s,min}^m}-\frac{q_{s,min}^m}{p_s^m}-1\Big).
\label{eq:xeq2}
\end{equation}
According to (\ref{eq:xeq}), $x$ is an increasing function of $E$. It means that with the increase in the ratio of the price to QoS, the regulatory organization forces ISPs to decrease the bandwidth dedicated to secondary users, in order to deal with ISP overcharging.
(\ref{eq:xeq}) is met as long as the parameter $E$ takes values such that $x$ falls within its valid range. We set (\ref{eq:xeq}) to one to find an upper bound for $E$,  denoted as $E_u$, which yields $E_u=0.7676$. In other words, the total ISP bandwidth is dedicated to the primary users for $E>E_u=0.7676$. 
Now, we   find the lower bound for $E$,  denoted as $E_u$. For this purpose, we set $x=0$ in  (\ref{eq:xeq}) and solve it for $E$, which results in:
\begin{equation}
E_l = \frac{1-2c^m + \sqrt{(1-2c^m)^2+12}}{6}.
\label{eq:El}
\end{equation}

\begin{figure}
	\centering
	\includegraphics[width=0.7\linewidth]{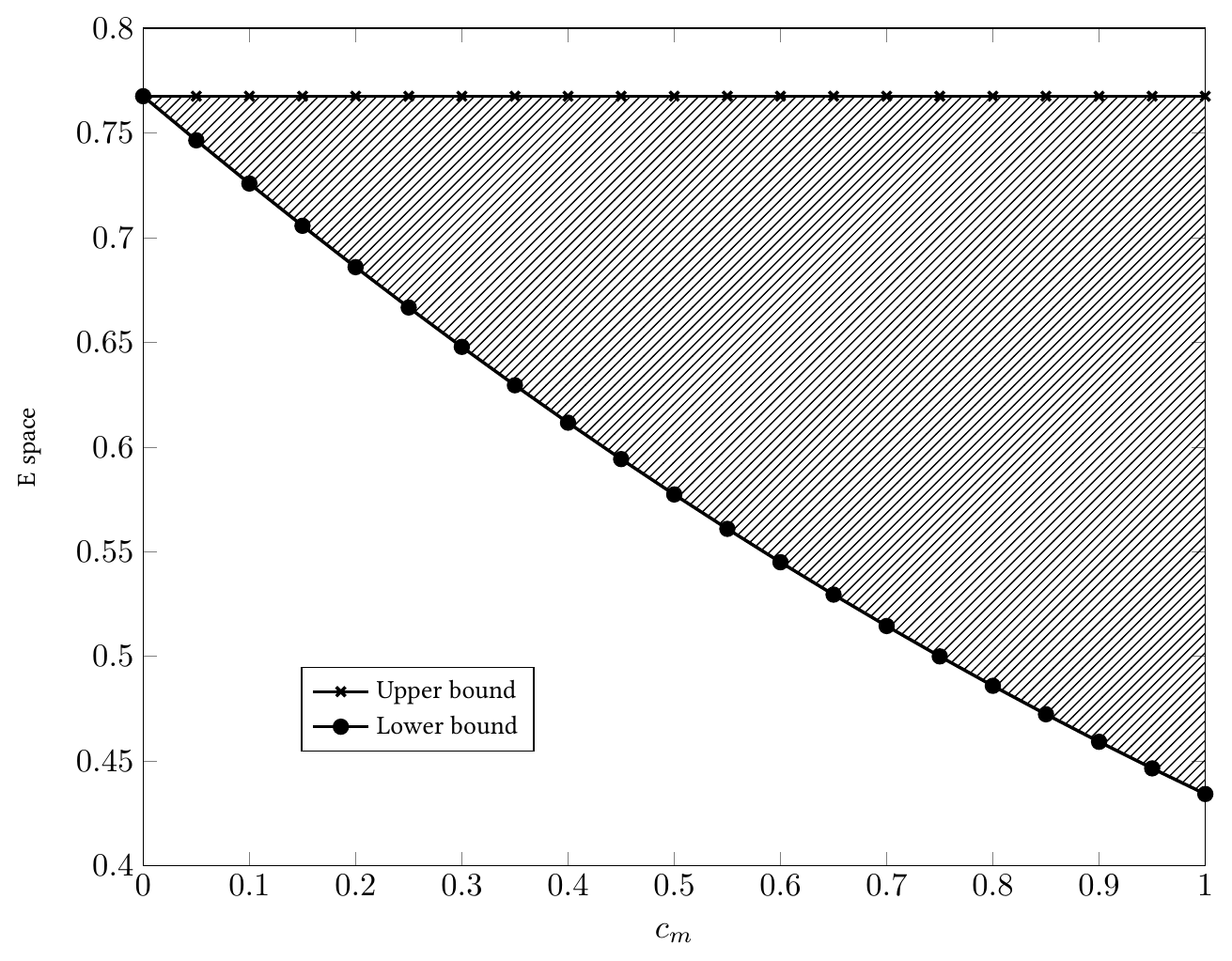}
\caption{Valid area for $E$ (hachured) against the normalized bandwidth.}
	\label{fig:bound}
\end{figure}

ISP dedicates its total bandwidth to the secondary users for $E<E_l$. 
  The valid $E$ is in the range of $E_l<E<E_u$. In Fig. \ref{fig:bound}, the valid  region  is shown  with respect to $c^m$. The ISP can increase its QoS with the increase in the perchased bandwidth per user, $c^m$. By increasing $c^m$, the valid region for $E$ expands. This fact is confirmed in Fig. \ref{fig:bound}. According to this figure, the lower bound of $E$ decreases with the increase in the normalized  bandwidth. 

\subsection{Primary Service Analysis}
\label{ssec:pa}

Similar to secondary users, primary users employ this service only when their utility function is always positive. Consecuently, setting $U_p^i\geq 0 $ in (\ref{eq:Utiltyp}), we have:
\begin{equation}
\theta_p^i \geq \frac{r_p^i p_p - \lambda I\Big(r_p^{i,k}-g(r_p^i,\alpha_k) \Big) }{ g(r_p^i,\alpha_k) }.
\label{eq:thetap}
\end{equation}

We aim to find the lower bound for $\theta_p^i$ conservatively. To this end, the right hand side of (\ref{eq:thetap}) needs to be maximized:
\begin{equation}
\frac{r_p^i p_p - \lambda I\Big(r_p^{i,k}-g(r_p^i,\alpha_k) \Big) }{ g(r_p^i,\alpha_k) }
\leq  
\frac{r_p^i p_p  }{ g(r_p^i,\alpha_k) }
\leq
\frac{ p_p  }{ g(r_p^i,\alpha_k) }
\leq
\frac{ p_p  }{ r_{min}}.
\label{eq:thetap2} 
\end{equation}

Therefore, the utility function of users whose type falls within the range of $[\frac{p_p}{r_{min}},1]$ is always a positive value and they use the primary services of the ISPs. Comparing the lower bound for the primary and secondary users,
  it can be deduced that the determining parameter in both of them is the price and minimum QoS that ISP provides for the users of each specific service.
By assuming a uniform distribution for the type of the primary users, their average utility can be found as below:
\begin{equation}
\overline{U}_p^i = \int_{\frac{p_p}{r_{min}}}^{1} U_p^i \mathrm{d}\theta_p^i = \frac{1}{2} \Big(1-(\frac{p_p}{r_{min}})^2\Big) g(r_p^i,\alpha_k)+\bigg( \lambda I\Big(r_p^{i,k}-g(r_p^i,\alpha_k) \Big)-r_p^i p_p\bigg) (1-\frac{p_p}{r_{min}}).
\label{eq:Ua}
\end{equation}

In (\ref{eq:Ua}), the parameters of the function $g(r_p^i,\alpha_k)$ in each timeslot are estimated  to minimize the difference between the momentary rate of the user, $r_p^{i,k}$, and the value dedicated to him at the $k$'th timeslot. Therefore, we can say:
\begin{equation}
\min_{\omega,\gamma_{j}} F(\omega,\gamma)
\stackrel{\Delta}{=}\Bigg((x-r_{min}) \bigg(\frac{\sum_{k'=k-t}^{k-1}\omega_{k'} \mathbf{1}\Big(r_p^{i,k'}-g(r_p^i,\alpha_{k'})\Big)}{t}\bigg)^{\gamma}+r_{min}-r_p^{i,k}\Bigg)^2.
\label{eq:opt}
\end{equation}
To minimize the objective function of  (\ref{eq:opt}), the parameters $\omega$ and $\lambda$ update instantly with observing each $r_p^{i,k}$ sample by employing the steepest descent method. The formulas for updating these parameters are given below:
\begin{align}
\omega^+ &=\omega^--\rho \frac{\partial F(\omega,\gamma_{j})}{\partial \omega}\bigg\vert^- \,\,\mathrm{and}
\label{opt2} \\
\gamma_{j}^+ &=\gamma_{j}^--\rho \frac{\partial F(\omega,\gamma_{j})}{\partial \gamma_{j}}\bigg\vert^-,
\label{eq:opt3}
\end{align}
where \lq$-$\rq and \lq$+$\rq signs denote the parameters in the previous and next steps.

Here, we analyze the capability of the proposed method in estimating the network burst traffic for primary users. Burst traffic is usually modeled by the long tail distributions \cite{pastor2007evolution}. According to the model proposed in \cite{17} and regarding the fact that the momentary user's traffic is normalized to $[0,1]$, the beta distribution is used to model the burst traffic. In \cite{newman2003structure} and \cite{sen2004analyzing}, it is mentioned that there exists a long-range dependency (LRD) between the network traffic  at different moments. Therefore, we set the Hurst parameter of this traffic to 0.8 using an auto-regressive  filter of the order 10. The Hurst parameter shows the LRD of the network traffic and falls within the range of $[0.5,1]$. 
The closer this parameters to 0.5,  the less the interdependency among moments of traffic.


\begin{figure}
\centering
\includegraphics[width=0.7\linewidth]{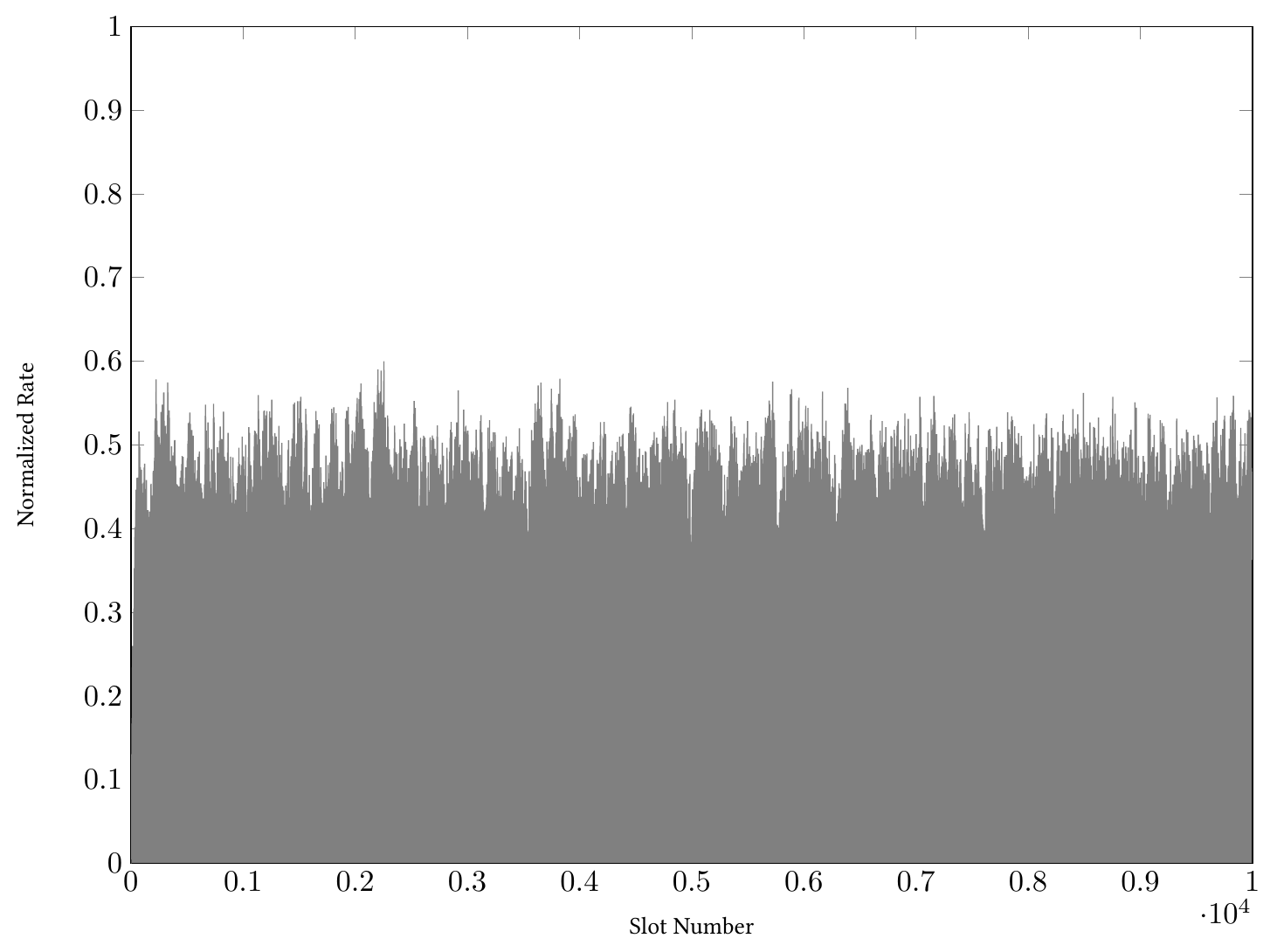}
\caption{A user's traffic simulation with $r_p^i=0.6$.}
\label{fig:tr2}
\end{figure}

\begin{figure}
\centering
\includegraphics[width=0.7\linewidth]{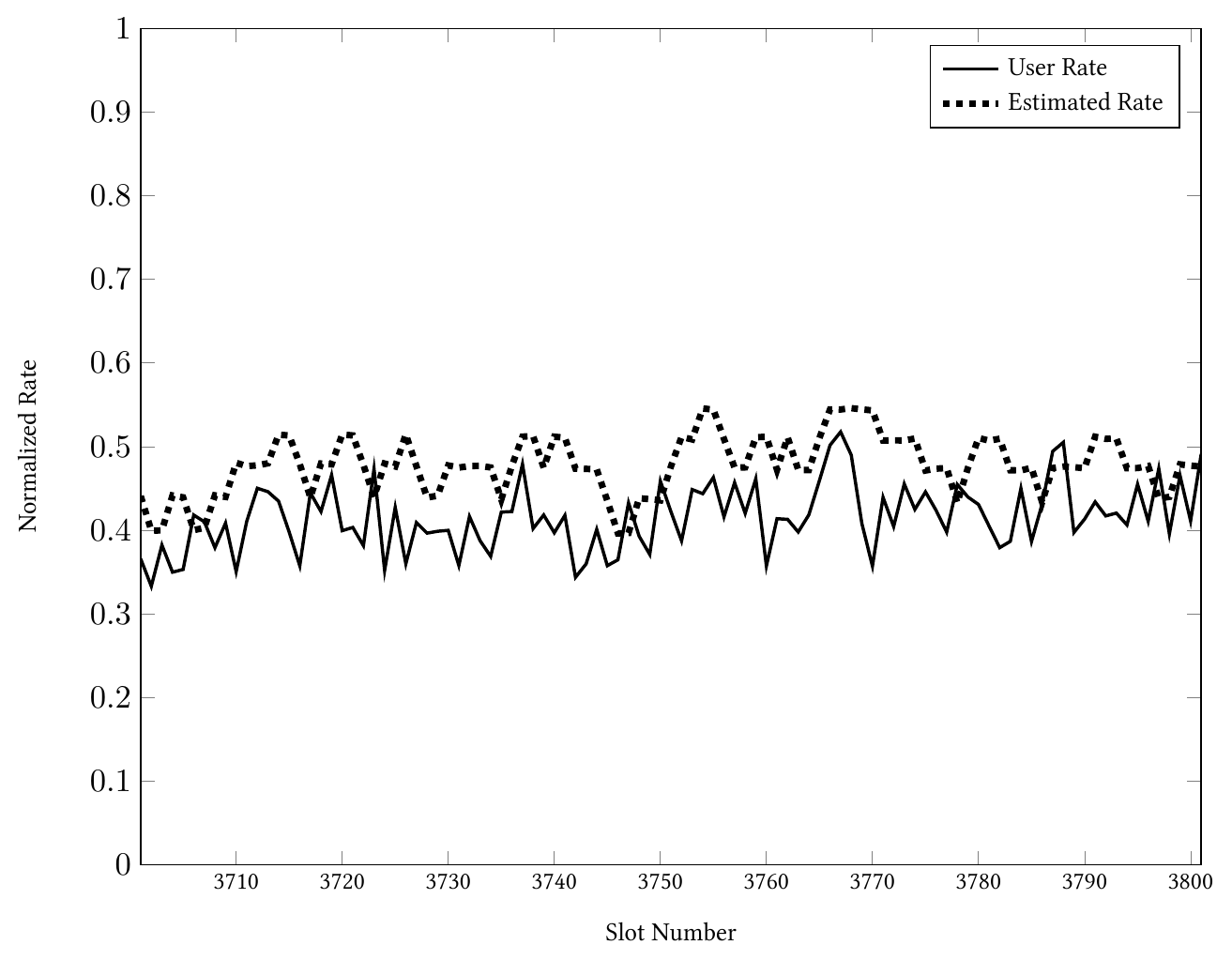}
\caption{Estimated traffic for the user
	 in Fig. \ref{fig:tr2} 
	in 100 timeslots.}
\label{fig:tr5}
\end{figure}

Fig. \ref{fig:tr2} 
presents the simulated traffic of some user in 10000 different timeslots, with the purchased bandwidth equal to $r_p^i = 0.6$, and the parameters of the beta distribution equal to $\alpha=3$ and $\beta=4$. Since the beta parameter takes values from $[0,1]$ range, it is multiplied by $r_p^i$ to find values from $[0,r_p^i]$. Fig. \ref{fig:tr5} shows the estimated and real rates in timeslots 3700 to 3800. The ISP has managed to provide the user with his demanded bandwidth in all timeslots using this method, without dedicating the total bandwidth bought by the user to him. Therefore, with appropriate management, the ISP can dedicate the remaining bandwidth to secondary users to increase their QoS. It can be observed in Fig. \ref{fig:tr5} that the estimated rate remains almost constant within timeslots that the user's rate does not vary greatly, due to the saturation of $g(r_p^i,\alpha_k)$ function.

\subsection{Effect of the Parameter of the Beta Distribution}
\label{ssec:beta}
\begin{figure}
\centering
\includegraphics[width=0.7\linewidth]{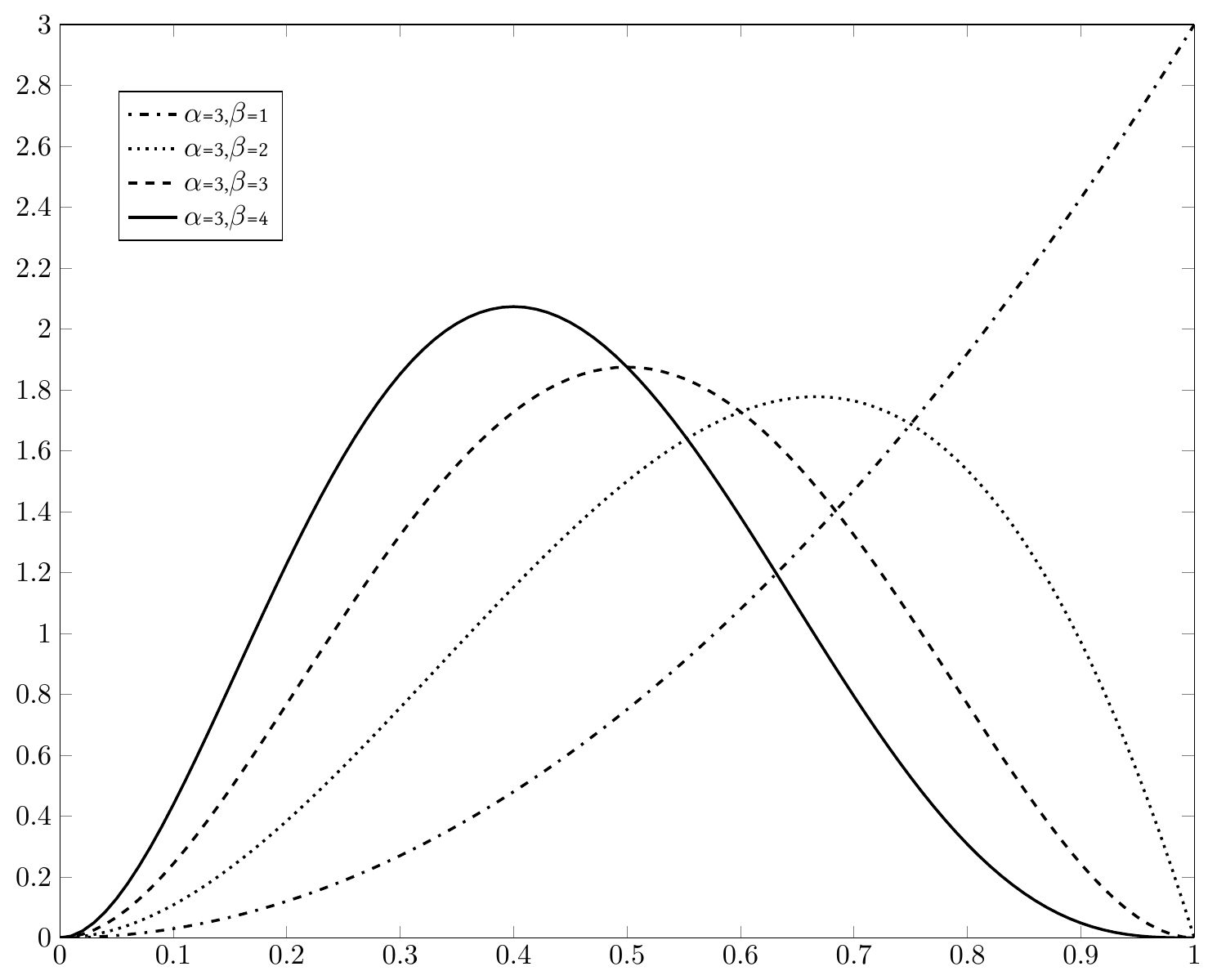}
\caption{Beta distribution of different parameters.}
\label{fig:beta}
\end{figure}

\begin{figure}
\centering
\includegraphics[width=0.7\linewidth]{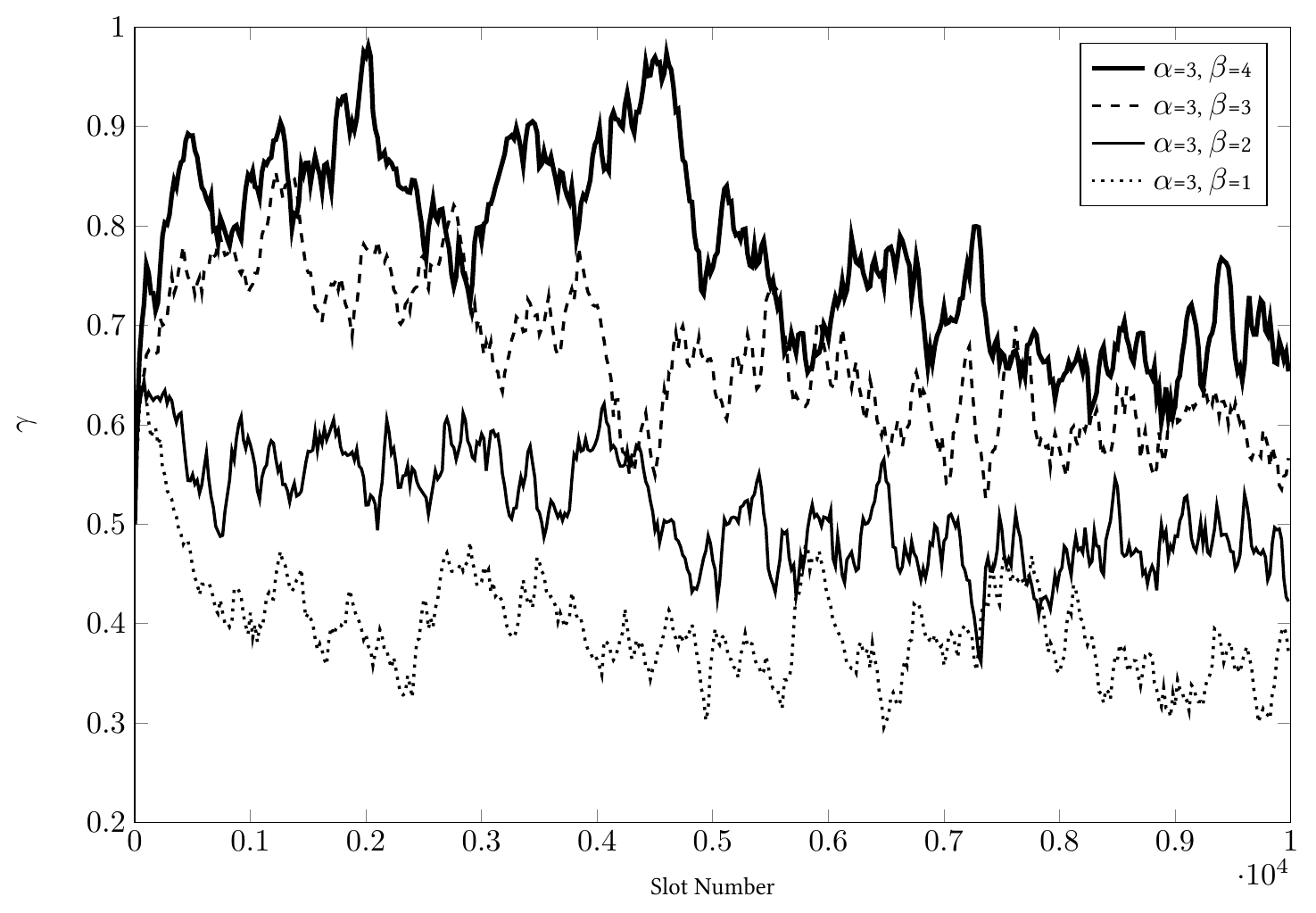}
\caption{Training the gamma parameter in $g(.)$ estimator against the changes in the parameters of the beta distribution.}
\label{fig:gamma}
\end{figure}

The beta distribution for some different parameters is given in Fig. \ref{fig:beta}. This distribution can model various scenarios of the user's application with different parameters. The user who maximally uses his purchased bandwidth can be modeled by parameters $\alpha=3$ and $\beta=1$, while the parameters $\alpha=3$ and $\beta=4$ model a user who does not fully use his purchased bandwidth. Fig. \ref{fig:gamma} demonstrates the training of the $\gamma$ parameter in estimating function $g(\cdot)$ against changes in the beta distribution parameter. When $\alpha=3$ and $\beta=1$, the user tends to exploites rates close to the purchased bandwidth. Thus, the $\gamma$ in the estimator must have a small value to properly work for  high rates. The fact in Fig. \ref{fig:gx} that $g(\cdot)$ function becomes more saturated in the less $\lambda$ values confirms this issue. As observed in Fig. \ref{fig:gamma}, the relative value of the $\gamma$ parameter increases with the increase in   parameter $\beta$. This fact is consistent with the interpretation given for the $\gamma$ parameter.

\section{Conclusion}
\label{sec:cncl}
In this research, a novel analytic model is proposed for more efficient services management  of the ISPs  which are in the same FCP. According to this model, the available bandwidth of the ISPs is divided into two parts: one for primary users and the other for secondary ones. This division is performed to maximize the benefit of the users. It is shown in section \ref{ssec:sa} that if the ratio between the price proposed by the ISP and the minimum presented QoS is greater than 0.7676, the ISP must refrain from supplying the secondary service in order to deal with ISP overcharging. On the other hand, for the ISP to maintain its own benefits, this ratio should not fall below a specific threshold. A reasonable range for this ratio is shown in Fig. \ref{fig:bound}.        
Since primary users do not always use the total bandwidth sold to them, consistent assignment of the total purchased bandwidth causes the resource loss. In order to better manage the available resources, ISP estimates the momentary rate of the primary users in each timeslot, and determines the bandwidth delivered to them based on this estimation. On the other hand, if ISP fails at providing the primary user with his required bandwidth, it must pay him a penalty proportional to the amount of the missing bandwidth.
  With this appropriate management, ISP can use the remaining bandwidth to increase the minimum QoS of secondary users. The proposed method would be more successful in optimally devoting the resources available to the network, compared with that proposed in \cite{2}, which is concentrated on dividing the bandwidth between the two types of service.  
  As the future work, we will consider more complicated and realistic utility function for the users to improve the proposed model. Moreover, it will be tried to propose a neural network approach to estimate the primary user’s rate in order to increase the QoS of both type users.

\bibliographystyle{IEEEtran}

\begin{thebibliography}{10}
	\providecommand{\url}[1]{#1}
	\csname url@samestyle\endcsname
	\providecommand{\newblock}{\relax}
	\providecommand{\bibinfo}[2]{#2}
	\providecommand{\BIBentrySTDinterwordspacing}{\spaceskip=0pt\relax}
	\providecommand{\BIBentryALTinterwordstretchfactor}{4}
	\providecommand{\BIBentryALTinterwordspacing}{\spaceskip=\fontdimen2\font plus
		\BIBentryALTinterwordstretchfactor\fontdimen3\font minus
		\fontdimen4\font\relax}
	\providecommand{\BIBforeignlanguage}[2]{{%
			\expandafter\ifx\csname l@#1\endcsname\relax
			\typeout{** WARNING: IEEEtran.bst: No hyphenation pattern has been}%
			\typeout{** loaded for the language `#1'. Using the pattern for}%
			\typeout{** the default language instead.}%
			\else
			\language=\csname l@#1\endcsname
			\fi
			#2}}
	\providecommand{\BIBdecl}{\relax}
	\BIBdecl
	
	\bibitem{1}
	P.~Dube and R.~Jain, ``Queueing game models for differentiated services,'' in
	\emph{Game Theory for Networks, 2009. GameNets '09. International Conference
		on}, May 2009, pp. 523--532.
	
	\bibitem{2}
	N.~Shetty, G.~Schwartz, and J.~Walrand, ``Internet {Q}o{S} and regulations,''
	\emph{IEEE/ACM Trans. Netw.}, vol.~18, no.~6, pp. 1725--1737, Dec. 2010.
	
	\bibitem{3}
	Comcast., "Comcast Powerboost," [Online]. Available:
	http://www.comcast.net/powerboost/.
	
	\bibitem{4}
	T.~Wu, ``Network neutrality, broadband discrimination,'' \emph{Journal of
		Telecommunications and high Technology law}, vol.~2, p. 141, 2003.
	
	\bibitem{5}
	J.~Krämer, L.~Wiewiorra, and C.~Weinhardt, ``Net neutrality: A progress
	report,'' \emph{Telecommunications Policy}, vol.~37, no.~9, pp. 794 -- 813,
	2013.
	
	\bibitem{6}
	M.~Mandjes, ``Pricing strategies under heterogeneous service requirements,''
	\emph{Computer Networks}, vol.~42, no.~2, pp. 231 -- 249, 2003.
	
	\bibitem{Etkin_2007}
	R.~Etkin, A.~Parekh, and D.~Tse, ``Spectrum sharing for unlicensed bands,''
	\emph{{IEEE} J. Select. Areas Commun.}, vol.~25, no.~3, pp. 517--528, apr
	2007.
	
	\bibitem{9}
	A.~Odlyzko, ``Paris metro pricing for the internet,'' in \emph{Proceedings of
		the 1st ACM Conference on Electronic Commerce}, ser. EC '99.\hskip 1em plus
	0.5em minus 0.4em\relax ACM, 1999, pp. 140--147.
	
	\bibitem{Shetty_2009}
	N.~Shetty, S.~Parekh, and J.~Walrand, ``Economics of femtocells,'' in
	\emph{{GLOBECOM} - {IEEE} Global Telecommunications Conference}, nov 2009.
	
	\bibitem{Ren_2012}
	S.~Ren and M.~van~der Schaar, ``Data demand dynamics in wireless communications
	markets,'' \emph{{IEEE} Transactions on Signal Processing}, vol.~60, no.~4,
	pp. 1986--2000, apr 2012.
	
	\bibitem{Lingjie_Duan_2014}
	L.~Duan, L.~Gao, and J.~Huang, ``Cooperative spectrum sharing: A contract-based
	approach,'' \emph{{IEEE} Transactions on Mobile Computing}, vol.~13, no.~1,
	pp. 174--187, jan 2014.
	
	\bibitem{Gao_2013}
	L.~Gao, J.~Huang, Y.-J. Chen, and B.~Shou, ``An integrated contract and auction
	design for secondary spectrum trading,'' \emph{{IEEE} J. Select. Areas
		Commun.}, vol.~31, no.~3, pp. 581--592, mar 2013.
	
	\bibitem{Shi_2014}
	Z.~Shi, K.~H. Li, T.~Tan, and K.~C. Teh, ``Energy efficient cognitive radio
	network based on multiband sensing and spectrum sharing,'' \emph{{IET}
		Communications}, vol.~8, no.~9, pp. 1499--1507, jun 2014.
	
	\bibitem{Sharma_2016}
	P.~K. Sharma and P.~K. Upadhyay, ``Cooperative spectrum sharing in two-way
	multi-user multi-relay networks,'' \emph{{IET} Communications}, vol.~10,
	no.~1, pp. 111--121, jan 2016.
	
	\bibitem{10}
	R.~Gibbens, R.~Mason, and R.~Steinberg, ``Internet service classes under
	competition,'' \emph{Selected Areas in Communications, IEEE Journal on},
	vol.~18, no.~12, pp. 2490--2498, Dec 2000.
	
	\bibitem{11}
	M.~de~Marin~de Montmarin, ``A result similar to the odlyzko's "paris metro
	pricing",'' \emph{Applied Economics}, vol.~38, no.~15, pp. 1821--1824, 2006.
	
	\bibitem{12}
	B.~E. Hermalin and M.~L. Katz, ``The economics of product-line restrictions
	with an application to the network neutrality debate,'' \emph{Information
		Economics and Policy}, vol.~19, no.~2, pp. 215 -- 248, 2007.
	
	\bibitem{13}
	N.~Economides and B.~E. Hermalin, ``The economics of network neutrality,''
	\emph{The RAND Journal of Economics}, vol.~43, no.~4, pp. 602--629, 2012.
	
	\bibitem{pastor2007evolution}
	R.~Pastor-Satorras and A.~Vespignani, \emph{Evolution and structure of the
		Internet: A statistical physics approach}.\hskip 1em plus 0.5em minus
	0.4em\relax Cambridge University Press, 2007.
	
	\bibitem{17}
	X.~Jian, X.~Zeng, Y.~Jia, L.~Zhang, and Y.~He, ``Beta/{M}/1 model for machine
	type communication,'' \emph{Communications Letters, IEEE}, vol.~17, no.~3,
	pp. 584--587, March 2013.
	
	\bibitem{newman2003structure}
	M.~E. Newman, ``The structure and function of complex networks,'' \emph{SIAM
		review}, vol.~45, no.~2, pp. 167--256, 2003.
	
	\bibitem{sen2004analyzing}
	S.~Sen and J.~Wang, ``Analyzing peer-to-peer traffic across large networks,''
	\emph{IEEE/ACM Transactions on Networking (ToN)}, vol.~12, no.~2, pp.
	219--232, 2004.
	
\end{thebibliography}

\end{document}